\documentclass[prd,tightenlines,twoside,secnumarabic,onecolumn,floatfix,nofootinbib,showpacs,11pt]{revtex4-1}

\pdfoutput=1
\usepackage[english]{babel}
\usepackage{calligra}
\usepackage{epsfig}
\usepackage{float}
\usepackage{multirow}
\usepackage{epstopdf}
\usepackage{amsmath,amssymb,bm,slashed}
\usepackage{mathtools}
\usepackage{graphicx}
\usepackage[sort&compress]{natbib}
\usepackage{xcolor}
\usepackage[normalem]{ulem}
\usepackage{hyperref}
\usepackage{cleveref}
\usepackage{subfigure} 
\definecolor{red}{rgb}{1.0, 0, 0}
\usepackage{slashed}
\usepackage{ctable}
\usepackage[utf8]{inputenc}

\allowdisplaybreaks

\bibpunct{[}{]}{,}{n}{}{,}

\definecolor{seagreen}{rgb}{0.180392,0.545098,0.341176}

\begin{document}

\title{LHC diphoton excess from colorful resonances}

\author{Jia Liu $^{1}$}
 \email[Email: ]{ liuj@uni-mainz.de}
\author{Xiao-Ping Wang $^{1}$ }
 \email[Email: ]{ xiaowang@uni-mainz.de}
\author{Wei Xue $^{2}$}
 \email[Email: ]{ weixue@mit.edu}

\affiliation{$^{1}$PRISMA Cluster of Excellence and Mainz Institute for Theoretical Physics,
Johannes Gutenberg University, 55099 Mainz, Germany\\ $^{2}$ Center for Theoretical Physics, Massachusetts Institute of Technology,
Cambridge, MA 02139, USA}

\date{\today} 


\begin{abstract}

Motivated by the possible diphoton excess around $750~\rm{GeV}$ observed by ATLAS and CMS at $13~\rm{TeV}$, we consider a
coloron model from $\rm{SU}(3)_1 \times \rm{SU}(3)_2$ spontaneously breaking to the Standard Model $\rm{SU}(3)_C$.
A colored massive vector boson is resonantly produced by $q \bar q $ in proton collision, followed by a colored scalar cascade decay.
This process gives two photons and one jet in the final states. And the kinetic edge of the two photons can be an interpretation of the diphoton excess, while satisfying
the dijet, $\rm{t}\bar{t}$, jet+photon resonance constraints. In this model,
due to the large mass of vector resonance, the parton luminosity function
ratio between $13~\rm{TeV}$ and $8~\rm{TeV}$ can be quite large.
Therefore, the diphoton excess has not been observed at $8~\rm{TeV}$ search.
On the other hand, having all the new particles color-charged around $\rm{TeV}$,
this model predicts new signals at the LHC, which can be validated soon.

\end{abstract}

\begin{flushright}
MITP/15-12X
\end{flushright}

\maketitle
\tableofcontents

\section{Introduction}
\label{sec:intro}

ATLAS and CMS collaborations present their RUN 2 results on inclusive diphoton search \cite{ATLAS-CONF-2015-081,
CMS-PAS-EXO-15-004} at
$\sqrt{s} = 13 \rm{TeV}$. Both of the experiments reveal an excess in the diphoton $m_{\gamma\gamma}$ spectrum
near $750~\rm{GeV}$. Having $3.2~\rm{fb^{-1}}$ data, ATLAS gives 3.9 (2.3) $\sigma$ local (global) significance for broad
resonance search and 3.6 (2.0) $\sigma$ local (global) significance for narrow width
approximation. CMS have 2.6 $\rm{fb}^{-1}$ data and gives $2.6$ $\sigma$ local significance.
Although this excess could be statistical fluctuations, it motivates many new physics explanations \cite{Mambrini:2015wyu,Harigaya:2015ezk,
Backovic:2015fnp,Nakai:2015ptz,Buttazzo:2015txu,Franceschini:2015kwy,DiChiara:2015vdm,Angelescu:2015uiz,Knapen:2015dap,Pilaftsis:2015ycr,
Ellis:2015oso,Bellazzini:2015nxw,Gupta:2015zzs,Molinaro:2015cwg,Higaki:2015jag,McDermott:2015sck,Low:2015qep,Petersson:2015mkr,Dutta:2015wqh,
Cao:2015pto,Matsuzaki:2015che,Kobakhidze:2015ldh,Cox:2015ckc,Ahmed:2015uqt,Agrawal:2015dbf,Martinez:2015kmn,Becirevic:2015fmu,No:2015bsn,
Demidov:2015zqn,Chao:2015ttq,Fichet:2015vvy,Curtin:2015jcv,Bian:2015kjt,Chakrabortty:2015hff,Csaki:2015vek,Falkowski:2015swt,Aloni:2015mxa,
Bai:2015nbs,Gabrielli:2015dhk,Benbrik:2015fyz,Kim:2015ron,Alves:2015jgx,Megias:2015ory,Carpenter:2015ucu,Bernon:2015abk,
Antipin:2015kgh,Wang:2015kuj,Cao:2015twy,Huang:2015evq,Liao:2015tow,Heckman:2015kqk,Dhuria:2015ufo,Bi:2015uqd,Kim:2015ksf,Berthier:2015vbb,
Cho:2015nxy,Cline:2015msi,Bauer:2015boy,Chala:2015cev,Kulkarni:2015gzu,Barducci:2015gtd,Boucenna:2015pav,Murphy:2015kag,Hernandez:2015ywg,
Dey:2015bur,Pelaggi:2015knk,deBlas:2015hlv,Belyaev:2015hgo,Dev:2015isx,huang_dec23,Moretti_dec23,Patel_dec23,Badziak_dec23,Chakraborty_dec23,
Cao_dec23,Altmannshofer_dec23,Cvetic_dec23,Gu_dec23}.

\begin{figure*}
    \includegraphics[width=0.4\textwidth]{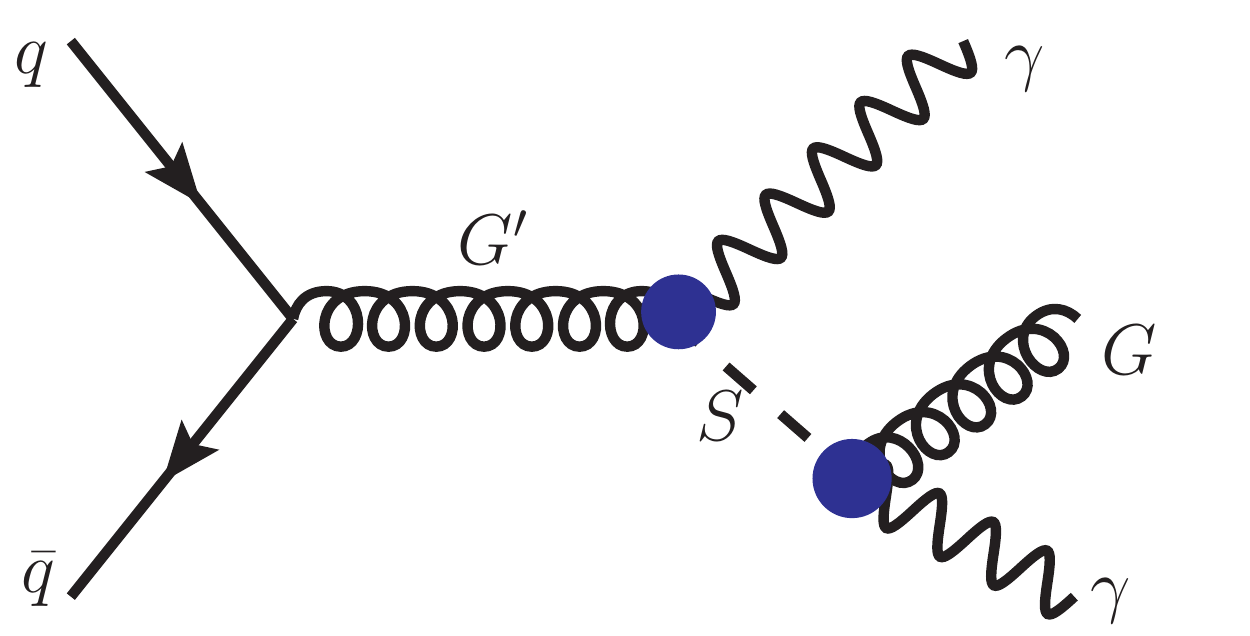}
  \caption{The Feynman diagram for resonance production of $G'$, which subsequently
  cascade decay to $S+\gamma \to G+\gamma+\gamma$. The blue blobs represent the fermion $\Psi$ loops.
  }
  \label{fig:feynman-Gp-resonance}
\end{figure*}

Due to Laudau-Yang theorem~\cite{Yang:1950rg, tagkey1965471}, the excess can be explained by $750~\rm{GeV}$ spin-0 particle produced resonantly, and meanwhile it needs other
particles introduced to have the right coupling to proton proton and to photon photon. Most of the models interpret the excess by introducing
a singlet scalar, while in our paper, we use particles with the Standard Model (SM) $\rm{SU}(3)_C$ color charge to produce the diphoton resonance.
We consider
a coloron model, $\rm{SU}(3)_1 \times \rm{SU}(3)_2 \rightarrow \rm{SU}(3)_C$ to give an alternative way to reproduce the excess in the diphoton spectrum.
The cascade decay of heavier colored particle to lighter colored particle, can produce
the kinetic edge in the diphoton invariant mass spectrum, see \cref{fig:feynman-Gp-resonance}.
This model is consistent with the collider constraints, like dijet, $t \bar t$ and
jet$+\gamma$ resonance at both 8 (13) TeV and also safe from diphoton resonance search
at 8 TeV. The model also predicts unique event topology, different from the 750 GeV
scalar resonance, including new resonances and
different kinetic distribution of the final state particles.

The paper is organized as follows. In section \ref{sec:model}, we introduce the coloron model, where a heavy color particle is the
massive gauge field after symmetry breaking and a relative lighter extra scalar is color charged as well. The cascade decay of the
two particles produce two photons and one gluon in the final states, which generate a kinetic edge in the di-photon spectrum.
In section \ref{sec:constraints}, we discuss the LHC constraints on the model from
dijet, $t \bar t$, jet$+\gamma$ and diphoton resonance searches at 8 and 13 TeV.
We find that the parameter space which can explain the di-photon excess
can be consistent with those collider constraints.
In section \ref{sec:diphoton}, we study the signal diphoton invariant mass spectrum
and its other kinetic properties. We summarize our results in section \ref{sec:conclusions}.

\section{Model}
\label{sec:model}

In this section, we start with the coloron model, which is gauged under
$\text{SU}(3)_1\times \text{SU}(3)_2$ spontaneously breaking down to QCD
$\text{SU}(3)_C$~\cite{Frampton:1987dn, Bagger:1987fz, Hill:2002ap}.
The spontaneous breaking is induced by a bi-triplet scalar $\Phi$, a $3\times 3$ matrix charged under $\text{SU}(3)_1\times \text{SU}(3)_2$ as $\left(3,\bar{3}\right)$ representation.
Besides the gauge fields, heavy vector fermions and color-octet scalar and their interactions are considered here.
The colored heavy particles
are produced by and can decay to the Standard Model particles.

\subsection{spontaneous breaking down to $\rm{SU}(3)_{\rm C}$}
\label{sec:downtosu3C}

The Lagrangian for the kinetic term for $G_1$, $G_2$ and scalar $\Phi$ are given as
\begin{align}
\mathcal{L}^{\rm{kin}}=-\frac{1}{4}G^a_{1,\mu \nu }G_1^{a,\mu \nu }-\frac{1}{4}G^a_{2,\mu \nu }G_2^{a,\mu \nu }+\text{Tr}\left[\left(D_{\mu }\Phi \right){}^{\dagger }D^{\mu }\Phi \right],
\label{eq:gaugeInter}
\end{align}
where $G_1$ and $G_2$ are the field strength of $\text{SU}(3)_1$ and $\text{SU}(3)_2$ respectively. $a$ is the color index running through 1 to 8. The covariant derivative of $\Phi$ field is,
\begin{align}
D_{\mu }\Phi =\partial _{\mu }\Phi -i g_1G_{1, \mu }^a T^a\Phi +i g_2G_{2,\mu }^a T^a\Phi ,
\label{eq:phiDeriv}
\end{align}
where $g_i (i=1,2)$ is the gauge coupling of the $\text{SU}(3)_i$ gauge group and $G_{i, \mu}$
is the corresponding gauge field for $\text{SU}(3)_i$.
We use $G_{i}$ for both gauge field strength and gauge field itself. When
this notation is confusing, we add Lorentz indices $\mu\nu$ or $\mu$ to distinguish them.

We take the spontaneous breaking pattern of $\Phi$ to be $\langle \Phi \rangle =
  v_{\Phi }\text{I}/\sqrt{6}$, where I is the $3\times3$ identity matrix.
We can read out the vector
which gets mass after the spontaneous breaking from
\begin{align}
\left\langle { D_{\mu }\Phi} \right\rangle \rightarrow
i \left(- g_1G_{1\mu }^a+  g_2G_{2\mu }^a\right)  T^a \frac{v_{\Phi}}{\sqrt{6}},
\end{align}
which shows that the linear combination $- g_1G_{1\mu }^a+  g_2G_{2\mu }^a$ obtain
mass and we denote it as $G'$. The other orthogonal combination remain
massless and which is in fact the QCD gluon, denoted as $G$. The mixing matrices
for $G_1$ and $G_2$ are given by

\begin{align}
\left(
\begin{array}{c}
 G_1 \\
 G_2
\end{array}
\right)=\left(
\begin{array}{cc}
 \text{cos$\theta $}_g & -\text{sin$\theta $}_g \\
 \text{sin$\theta $}_g & \text{cos$\theta $}_g
\end{array}
\right)\left(
\begin{array}{c}
 G' \\
 G
\end{array}
\right),
\label{eq:mixing}
\end{align}
where the mixing angle $\theta _g$ is defined by
$\sin \theta_g \equiv -g_2/\sqrt{g_1^2+g_2^2}$ and $\cos \theta_g \equiv g_1/\sqrt{g_1^2+g_2^2}$.
The mass of the colored vector $G'$ is $m_{G'}^2=\left(g_1^2+g_2^2\right)v_{\Phi }^2/6$.

Considering the mixing between $G^\prime$ and $G$, we can get the
strong coupling of the gluon self-interaction term
\begin{align}
g_s \equiv  g_1g_2/\sqrt{g_1^2+g_2^2}.
\label{eq:gs}
\end{align}
The interactions between $G'$ and gluon $G$ are given as
\begin{align}
 \mathcal{L}^{\rm int}_{G'G}=& \frac{1}{2}g_s^2f^{\text{abc}}f^{\text{ade}}G'^{\text{$\mu $b}}\left\{G^{\text{$\nu $d}}\left(G'^c_{\nu} G_{\mu }^e+G_{\nu }^c G'^e_{\text{$\mu $}}\right)+
 G'^e_{\text{$\nu $}} G^{\text{$\nu $c}}G_{\mu }^d\right\} \nonumber \\
 & +g_s f^{\text{abc}}G'^a_{\text{$\mu $}}\left\{\left(\partial ^{\mu }G'^{\text{$\nu $b}}-\partial ^{\nu } G'^{\text{$\mu $b}}\right)G_{\nu }^c-G'^b_{\text{$\nu $}} \partial ^{\mu }G^{\text{$\nu $c}}\right\}.
 \label{eq:GpandGterm}
\end{align}

We can see that in \cref{eq:GpandGterm}, $G'$ always appears as quadratic term.
The reason is the kinetic term for $G_1$ and $G_2$ is symmetric under the operation
$1\leftrightarrow 2$. Under this operation, $G'$ get a minus sign, while $G$ does not change. Therefore, only quadratic $G'$ appears in \cref{eq:GpandGterm}. This prevent
$G'$ from decaying into gluons. After symmetry breaking, $G'$ can be considered as a matter field charged under $\rm{SU}(3)_{\rm C}$. In $G'$ kinetic term,
there is no linear term in $G'$ as well, for the same reason in \cref{eq:GpandGterm}.

\subsection{The interactions for extra fermions and scalars}
\label{sec:downtosu3C}

We assign the SM quarks $q_{\text{SM}}$ charged only under $\text{SU}(3)_2$ as fundamental
representation. Their $\text{SU}(2)_{\rm L}$ and $\text{U}(1)_{\rm Y}$ charges are the same as that in SM,
therefore we do not repeat them in the table. $q_{\text{SM}}$ is coupled to octet $G'$ through $G_{1,2}$ mixing, which is responsible for the resonance production of $G'$.
Here we also introduce a $\text{SU}(3)_1$ octet scalar $S$ without any other charge, and a heavy vector like fermion $\Psi$, charged only under
$\text{SU}(3)_1$ as fundamental representation and also under hypercharge $U(1)_{\rm Y}$.
It has a dimension 5 operator $S^a G_{1,\mu \nu }^a B_{\mu \nu }$, which will induce the interactions for $S  \gamma G^\prime$ and $S \gamma G$. The mass of $\Psi$ is
assumed to be high to avoid $G'$ and $S$ decaying to $\Psi$ pair.
The summary of the gauge charges for the particles are listed in \cref{tab:particles}.

\begin{table}
  \centering
  \begin{ruledtabular}
  \begin{tabular}{lcccc}
         & $\text{SU}(3)_1$ & $\text{SU}(3)_2$ & $\text{SU}(2)_L$ & $\text{U}(1)_Y$ \\
  $G_1$  & 8 & 1 & 1 & 0 \\
  $G_2$  & 1 & 8 & 1 & 0 \\
 $\Phi$  & 3 & $\bar 3$ & 1 & 0 \\
 $\Psi$  & 3 & 1 & 1 & $Q_Y^{\Psi}$ \\
 $S$     & 8 & 1 & 1 & 0 \\
 $q_{\text{SM}}$& 1 & 3 & $Q^q_L$ & $Q^q_Y$ \\
  \end{tabular}
  \end{ruledtabular}
  \caption{List of  particle contents and their gauge charges.
  $G_1$ and $G_2$ are the vector gauge bosons for $\text{SU}(3)_1$ and $\text{SU}(3)_2$
  respectively. $\Phi$ is a $\text{SU}(3)_1\times\text{SU}(3)_2$ bi-triplet, which is
  responsible for breaking $\text{SU}(3)_1\times\text{SU}(3)_2$ into $\text{SU}(3)_{\rm C}$.
  $\Psi$ is a heavy vector-like fermion and charged under $\text{SU}(3)_1$ and $\text{U}(1)_{\rm Y}$.
  The SM quarks $q_{\text{SM}}$ are charged under $\text{SU}(3)_2$, while the $\text{SU}(2)_{\rm L}$ and $\text{U}(1)_Y$ charges are the same as that in SM which we do not
  explicitly write down. $q_{\text{SM}}$ couples to octet $G'$ through $G_{1,2}$ mixing,
  which is responsible for the resonance production of $G'$.
  }
  \label{tab:particles}
\end{table}

We list the relevant dimension 4 gauge interactions for fermions below
\begin{align}
 \mathcal{L}_{4d,F}^{{\mathop{\rm int}} } & =  - ig_1 G_{1,\mu }^a \bar \Psi \gamma ^\mu  T^a \Psi  - ig_2 G_{2,\mu }^a \bar q_{\rm SM} \gamma ^\mu  T^a q_{\rm SM} \nonumber \\
  & =  - i\left( {g_1 \cos \theta _g G'^a_\mu  + g_s G_\mu ^a } \right)\bar \Psi \gamma ^\mu  T^a \Psi  - i\left( {g_2 \sin \theta_g G'^a_\mu  + g_s G_\mu^a } \right)\bar q_{\rm SM} \gamma ^\mu  T^a q_{\rm SM},
  \label{eq:L4dF}
\end{align}
 where in the second line we expand $G_{1,2}$ into their mass eigenstates $G^\prime$ and $G$.
For both $\Psi$ and $q_{\rm SM}$, the coupling to $G$ is $g_s$ due to they are fundamental
representation of $\rm{SU}(3)_{\rm C}$.
Since we assume $\Psi$ is very heavy,
the production of a $\Psi$ pair through $G$ is tiny. In our model, $\Psi$ can be
unstable and decay into SM particles to avoid cosmological limits
\footnote{If $Q_Y^{\Psi} = Q_Y^{u_R}$ where $u_R$ is
the SM right handed up quarks, $\Psi$ can decay via Yukawa term $\bar \Psi \Phi u_R$.
In \cref{eq:gaugeInter}, after $\Phi$ obtain its vev, $\Phi$ can decay into two $G'$
which is similar as SM Higgs decay to two Z boson.
In more detail, the $3\times 3$ matrix form of $\Phi$ can be written as $
\Phi  = \phi _0 I + \phi _8^a T^a$. \cref{eq:gaugeInter} contains operators like
$\phi _0 G'^a_\mu G'^a_\mu$ and $d_{abc} \phi _8^a G'^b_\mu G'^c_\mu$, which
can mediate
the decay. Since $G'$ can further decay to SM quarks, $\Psi$ is unstable.
 }.

The relevant dimension 4 operators for scalar $S$ are given by
\begin{align}
 \mathcal{L}_{\rm 4d,S}^{{\mathop{\rm int}} } & \supset \left( {D_\mu  S} \right)_a^\dag  \left( {D_\mu  S} \right)_a  + \lambda S^a \bar \Psi T^a \Psi,  \label{eq:L4dS}  \\
 \left( {D_\mu  S} \right)_a & = \alpha _\mu  S_a  - ig_1 f_{abc} G_{1,\mu }^b S^c  = \alpha _\mu  S_a  - if_{abc} \left( {\cos \theta _g g_1 G'^b_\mu   + g_s G_\mu ^b } \right)S^c,
 \label{eq:derivativeS}
\end{align}
where the color octet $S$ is charged under $\rm{SU}(3)_{\rm C}$, which can be pair produced
at the LHC. However, $S$ does not couple to SM quarks, because they are assigned
into different $\rm{SU}(3)$ groups respectively and also due to the chirality of SM quarks.

\subsection{The decay and production of $G'$ and $S$}
\label{sec:decayofGpandS}

We are interested in the mass hierarchy where $m_\Psi  ,m_\Phi   > m_{G'}  > m_S$.  How does $G'$ and $S$ decay in such hierarchy is important.
The Yukawa term of $S$ in \cref{eq:L4dS} can induce a dimension 5 operator,
\begin{align}
 \mathcal{L}_{5d}^{{\mathop{\rm int}} } & = k_S S^a G_{1,\mu \nu }^a B_{\mu \nu }  = k_S S^a \left( {\cos \theta _g G'^a_{\mu \nu }  - \sin \theta _g G_{\mu \nu }^a } \right)B_{\mu \nu } ,
 \label{eq:L5d}
 \end{align}
 with
 \begin{align}
 k_S  & = Q_Y^\Psi  e\frac{{\lambda g_1 }}{{32\pi ^2 m_\Psi  }}\tau \left( {1 + (1 - \tau )f(\tau )} \right),
\label{eq:ks}
\end{align}
where $\tau  = 4m_\Psi ^2 /m_S^2$ and loop function~\cite{Ellis:1975ap,
Shifman:1979eb}
\begin{align*}
  f(\tau) = \begin{cases}
           \arcsin^2(1/\sqrt{\tau})                  & \text{for $\tau \ge 1$} \\
           \frac{1}{2} \left[ \log\left( \frac{1 + \sqrt{1-\tau }}{1 - \sqrt{1-\tau}} \right)
                            - i\pi  \right]^2     & \text{for $\tau < 1$}
         \end{cases} \,.
\end{align*}
As $\Psi$ is heavier than $S$, the loop factor $\tau \left( {1 + (1 - \tau )f(\tau )} \right) = 2/3$ in the limit of $\tau\to\infty$.

In our mass setup, $S$ can only decay to $G+\gamma/Z$ via \cref{eq:L5d}. The corresponding decay widths of S are given by
\begin{align}
\Gamma _{S \to G+\gamma } & = \left( {k_S \sin \theta _g \cos \theta _W } \right)^2 \frac{{m_S^3 }}{{8\pi }}, \\
\Gamma _{S \to G + Z} & \approx \left( {k_S \sin \theta _g \sin \theta _W } \right)^2 \frac{{m_S^3 }}{{8\pi }}.
\end{align}
Finally, the branching ratios of $S$ decaying to $ G \gamma$ and $G Z$ equal to $\cos^2\theta_W = 77\%$ and $\sin^2\theta_W = 23\%$ respectively,
where $\theta_W$ is the Weak angle as same in SM. Due to significant
smaller branching ratio to $Z$ and high sensitivity for $\gamma$ detection, we will focus
on the decay channel $S \to G + \gamma$. But for high luminosity 13 TeV data,
the decay channel through $Z$ is definitely worth exploring.

\begin{figure}[t]
\begin{tabular}{cc}
    \includegraphics[width=0.48\textwidth]{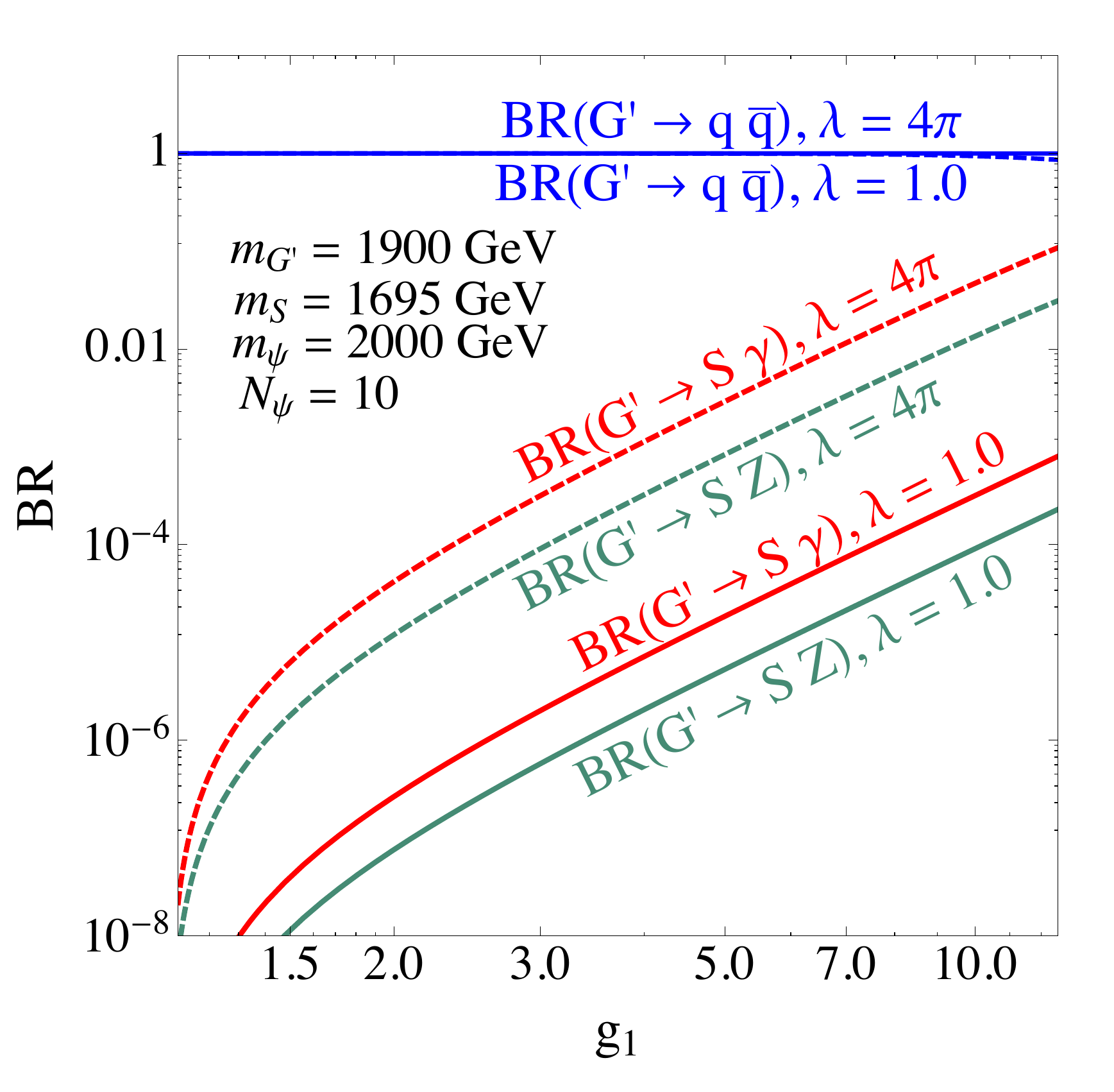}&
    \includegraphics[width=0.48\textwidth]{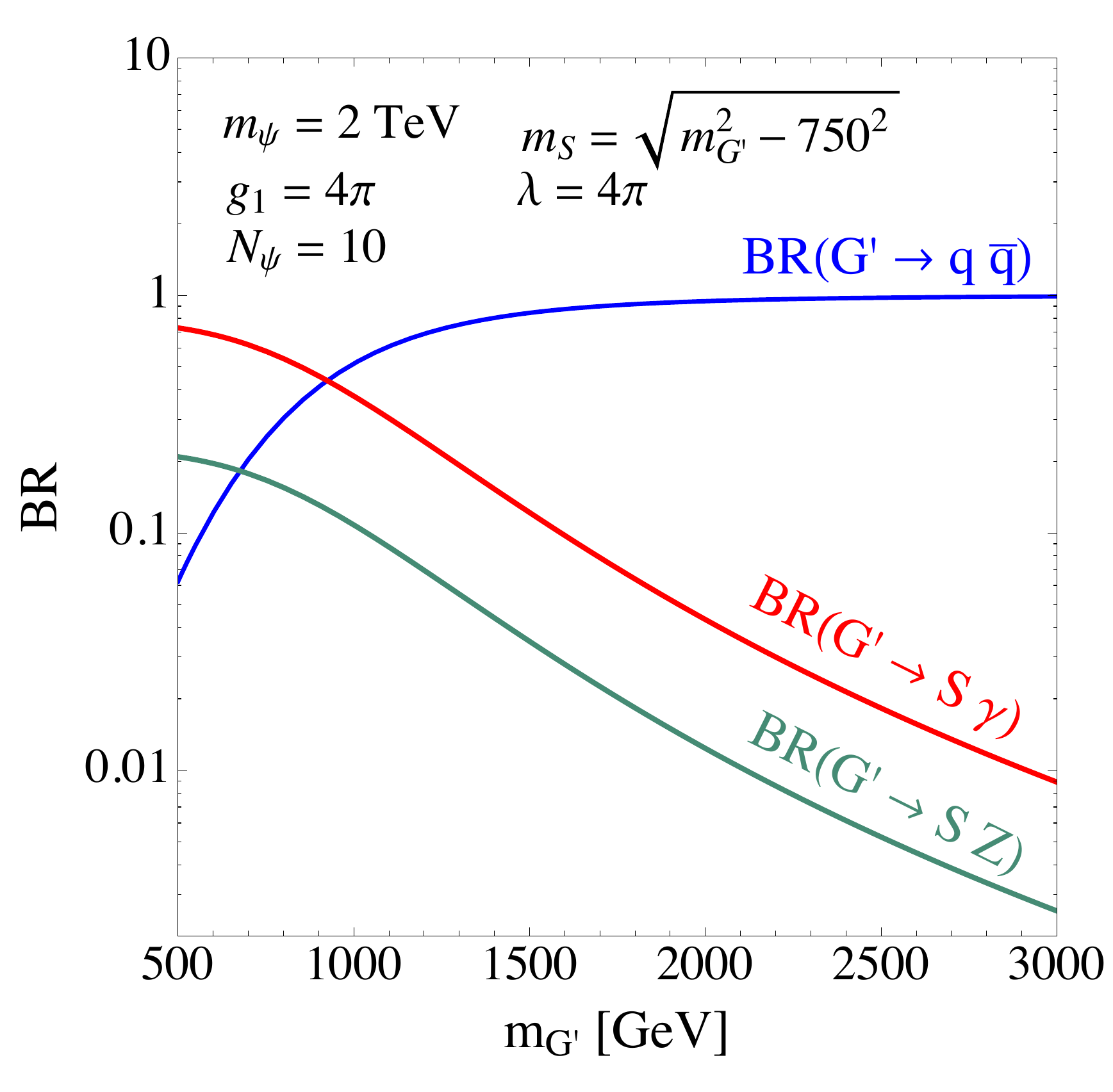}\\
    (a) & (b)
    \end{tabular}
  \caption{
  The $G' $ decay branching ratio for $\bar q q$ (including all six flavor), $S+\gamma$ and $S+Z$ channels. $N_\Psi $ is the copy of $\Psi$ which has
same hypercharge $Q_Y^{\Psi} =1$. (a) the $G'$ branching ratio as a function of $g_1$.
(b) the $G'$ branching ratio as a function of $m_{G'}$.
  }
  \label{fig:single-GP-decay}
\end{figure}

After mixing between $G^\prime$ and $G$, $G^\prime$ can decay to
a pair of SM quarks.  It also can decay to $S+\gamma/Z$ via loop operator in \cref{eq:L5d}. The formula of $G'$ decay widths are given by
\begin{align}
 \Gamma _{G' \to \bar qq} & = \frac{{\left( {g_2 \sin \theta _g } \right)^2 }}{{24\pi }}m_{G'} \left( {1 + 2\frac{{m_q^2 }}{{m_{G'}^2 }}} \right)\sqrt {1 - 4\frac{{m_q^2 }}{{m_{G'}^2 }}} \label{eq:Gpwidth1} ,\\
 \Gamma _{G' \to S + \gamma } & = \left( {k_S \cos \theta _g \cos \theta _W } \right)^2 \frac{{\left( {m_{G'}^2  - m_S^2 } \right)^3 }}{{24\pi m_{G'}^3 }}  \label{eq:Gpwidth2}, \\
\Gamma _{G' \to S + Z} & \approx \left( {k_S \cos \theta _g \sin \theta _W } \right)^2 \frac{{\left( {m_{G'}^2  - m_S^2 } \right)^3 }}{{24\pi m_{G'}^3 }}.
\label{eq:Gpwidth3}
\end{align}

We show the $G'$ decay branching ratio in \cref{fig:single-GP-decay}.
We generally need large $G'\to S+\gamma$ branching ratio to satisfy diphoton signal requirement
and avoid collider constraints at the same time.
Therefore, we assume a small mixing angle $\theta_g$, which
suggest that $g_1  \gg g_2 $. This will suppress
the decay width of $G'\to q \bar q$ by the coupling square
$(g_2^2 /\sqrt {g_1^2  + g_2^2 })^2$ .
To enhance the branching ratio of $G'\to S+\gamma$,
we use $g_1 = 4\pi$, $\lambda = 4\pi$ and large $N_\Psi $ copy of $\Psi$ which has
same hypercharge $Q_Y^{\Psi} =1$.

\begin{figure}[t]
    \includegraphics[width=0.48\textwidth]{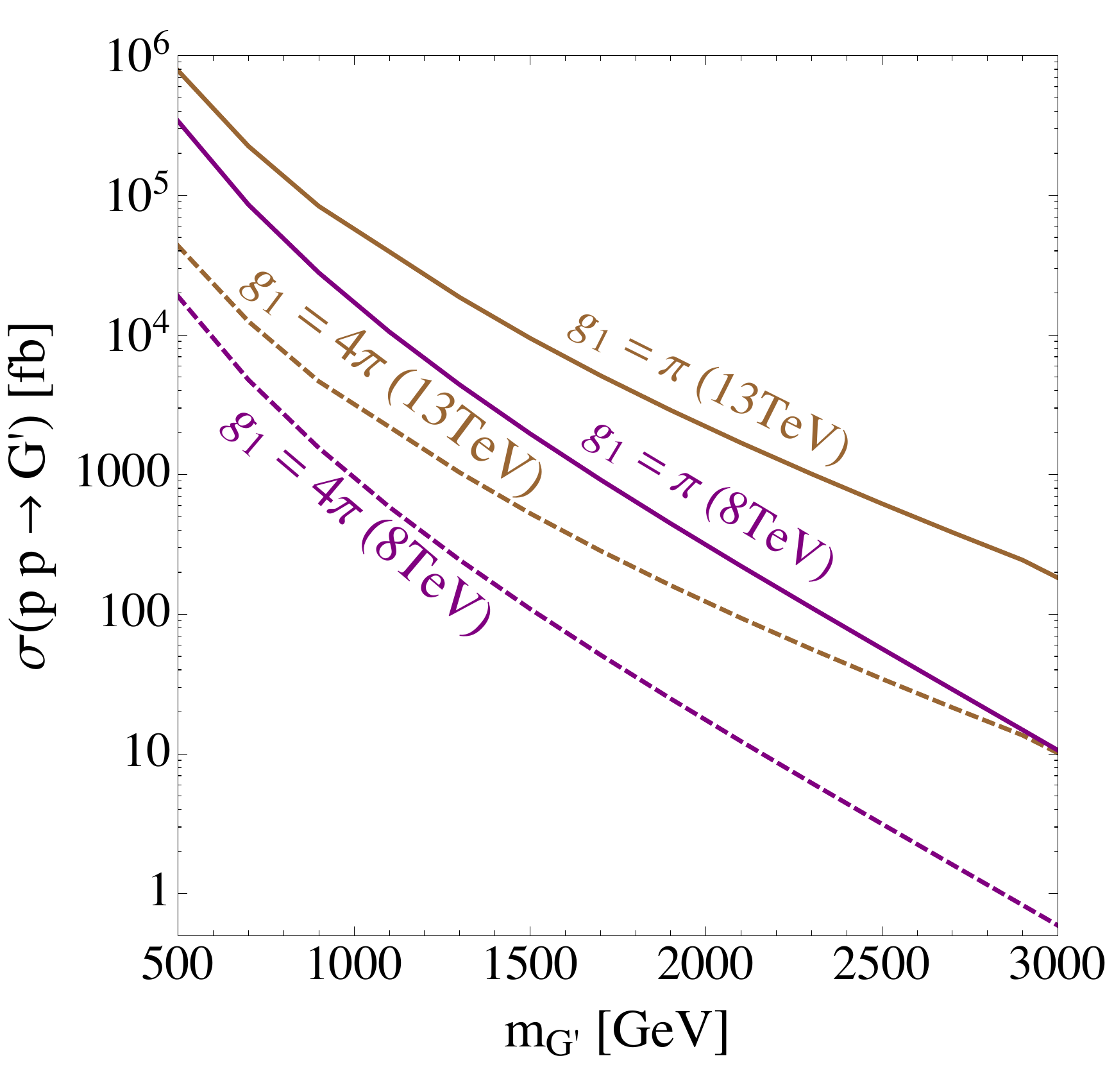}
    \includegraphics[width=0.48\textwidth]{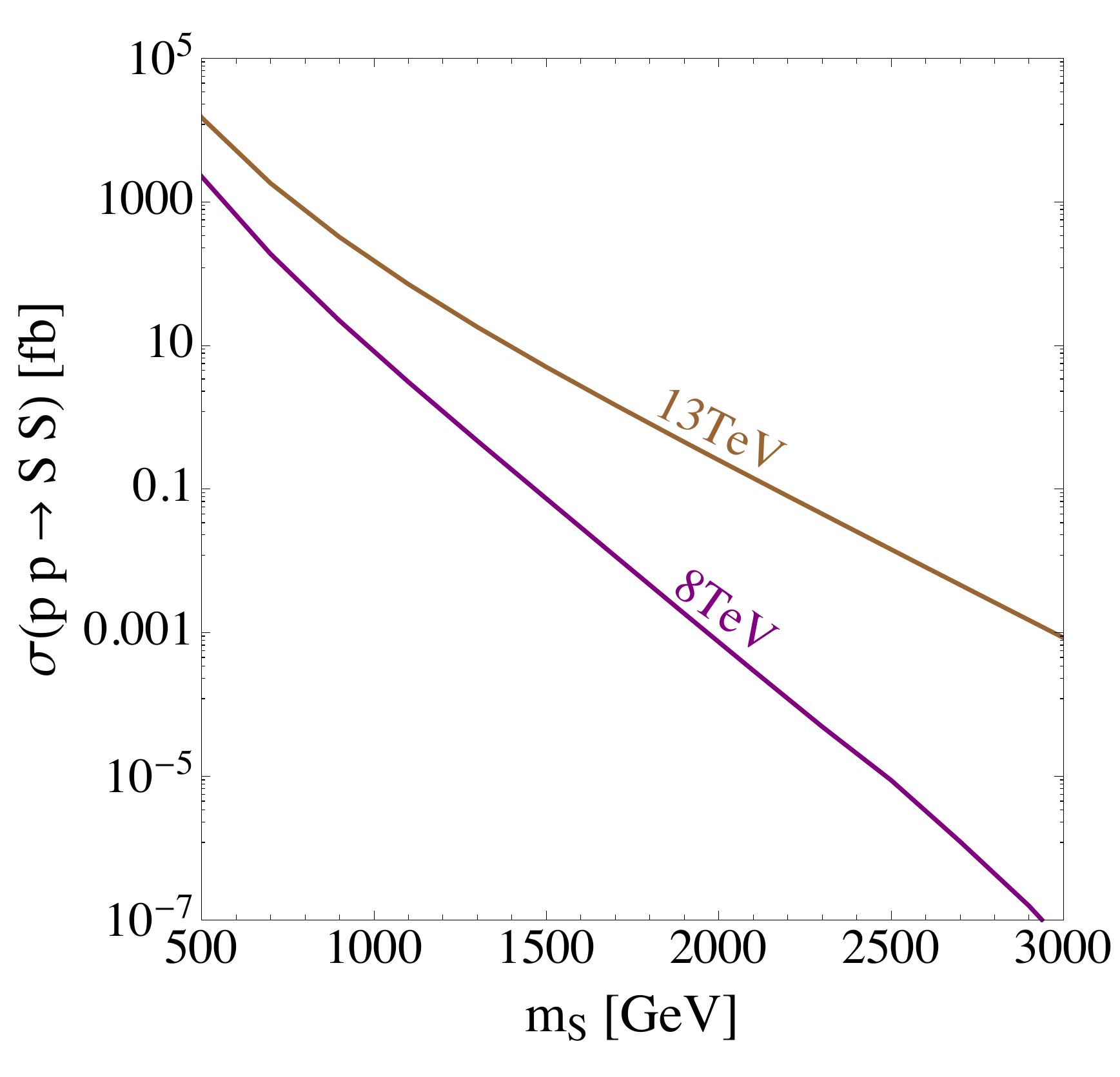}
  \caption{The single resonant production cross-section for $\bar q q \to G'$
   and the pair production cross-section for $p p \to S S$. Note that the $S$ pair production
   cross-section is only determined by strong interaction coupling $g_s$ and $m_S$.
  }
  \label{fig:single-GP-prod}
\end{figure}

In \cref{fig:single-GP-prod}, we show the single resonant
production $\bar q q \to G'$ cross-section at 8(13) TeV as a function of $m_{G'}$.
The calculation is done at tree level by \textsc{MadGraph~5~v2.3}~\cite{Alwall:2014hca},
with model implemented by \textsc{FeynRules~v2.3}~\cite{Alloul:2013bka}.
We do not apply K-factor from QCD correction because it is only about 1.2 for $q \bar q$
production~\cite{Franceschini:2015kwy}.
We also show cross-section for $S$ pair production for constraint purpose.

\section{Constraints for the model}
\label{sec:constraints}

In our model, the resonance production of single $G'$ will dominantly decay to
SM quarks. This process will be constrained by the ATLAS 8 (13) TeV dijet
invariant mass search \cite{Aad:2014aqa,ATLAS:2015nsi}. We choose their limits
for $W'$ hadronic decay and compare with our $G'$ resonance cross-section,
because they are both vector boson that the acceptance will be similar.
For 8 (13) TeV, the dijet constraint shows that the cross-section times acceptance
should be smaller than 60 (150) fb for $W'$ at 2 TeV. We can see at this mass, our
resonance cross-section of single $G'$ is already smaller than the constraint.
Including the acceptance will further weaken the constraints. Moreover, we should
multiply by $5/6$ from decay branching ratio of $G' \to \bar q q$,
since it only constraints five flavor of quarks without top.
If we conservatively assume that our signal acceptance is 1,
we conclude that $G'$ mass should be higher than 1450 (1250) GeV for 8 (13) TeV data.

The single resonance $G'$ can also decay to $\bar t t$ whose branching ratio is
$1/6$ of the total branching ratio of $G\prime\to q \bar q$. ATLAS has searched for $\bar t t$ resonance through lepton-plus-jets
topology at 8 TeV
\cite{Aad:2015fna}. Their constraint for a resonant $Z'$ which decay to $\bar t t$,
is smaller than 50 fb at $m_{Z'} \sim \rm{2TeV}$. This limit is comparable
to dijet search. After considering the branching ratio of $G' \to \bar t t$ ,
$\bar tt$ constraint is insensitive to the $G'\to t \bar t$ scenario.

Besides the constraints from $G'$, we should also consider the constraints
from $S$, which decay to $G+\gamma$ with branching ratio around $77\%$.
There are searches for jet-photon resonance from ATLAS at both
8 (13) TeV \cite{Aad:2013cva, Aad:2015ywd}. The $95\%$ CL limits on cross-section times acceptance are around 1 (10) fb at 8 (13) TeV respectively, for a
jet$+\gamma$ resonance at 2 TeV. This search can place a stringent constraint for
our model.

The $S$ pair could be produced through gluon fusion $GG \to SS$ with
strong coupling $g_s$, which is shown in \cref{eq:L4dS}. The pair production cross-section is given
in \cref{fig:single-GP-prod}.
The $S$ pair production cross-section decreases fast with increasing $m_S$, due to the two heavy $S$ in the final states.

Therefore, we should also consider single productions of $S$, which might have
larger cross-section. Fisrt there is no single production of $S$ from $G$ coupling as shown in \cref{eq:L4dS}. However, it can be produced from cascade decay of $G'$ and
associate production with a photon via coupling $G-\gamma-S$.

For the cascade decay, it depends on the single production cross-section of $G'$
and decay branching ratio $BR(G' \to S + \gamma)$.
$G'$ single production cross-section is determined by its quark coupling
$g_2 \sin\theta_g$, where we choose $g_1 = 4 \pi$  to fix the coupling
due to the diphoton signal requirement. The decay branching ratio of $G' \to S + \gamma$
depends on masses and couplings.

For the associate production, we have calculated its cross-section using the same parameter in $G'$ cascade by \textsc{MadGraph~5~v2.3}~\cite{Alwall:2014hca}.
We found that it is much smaller than $G'$ cascade. By comparing the couplings and
branching ratio formula, both of them are suppressed by the dimension 5 operator.
However, the dimension 5 coupling for associate production is further
suppressed by $\sin\theta_g$, see \cref{eq:L5d}.
Furthermore, it is a 2 to 2 production, while $G'$ cascade is a single resonant
production. As a result, $S$ associate production with photon does not provide
constraints nor can it provide diphoton events to the signal.

\begin{figure*}
    \includegraphics[width=0.5\textwidth]{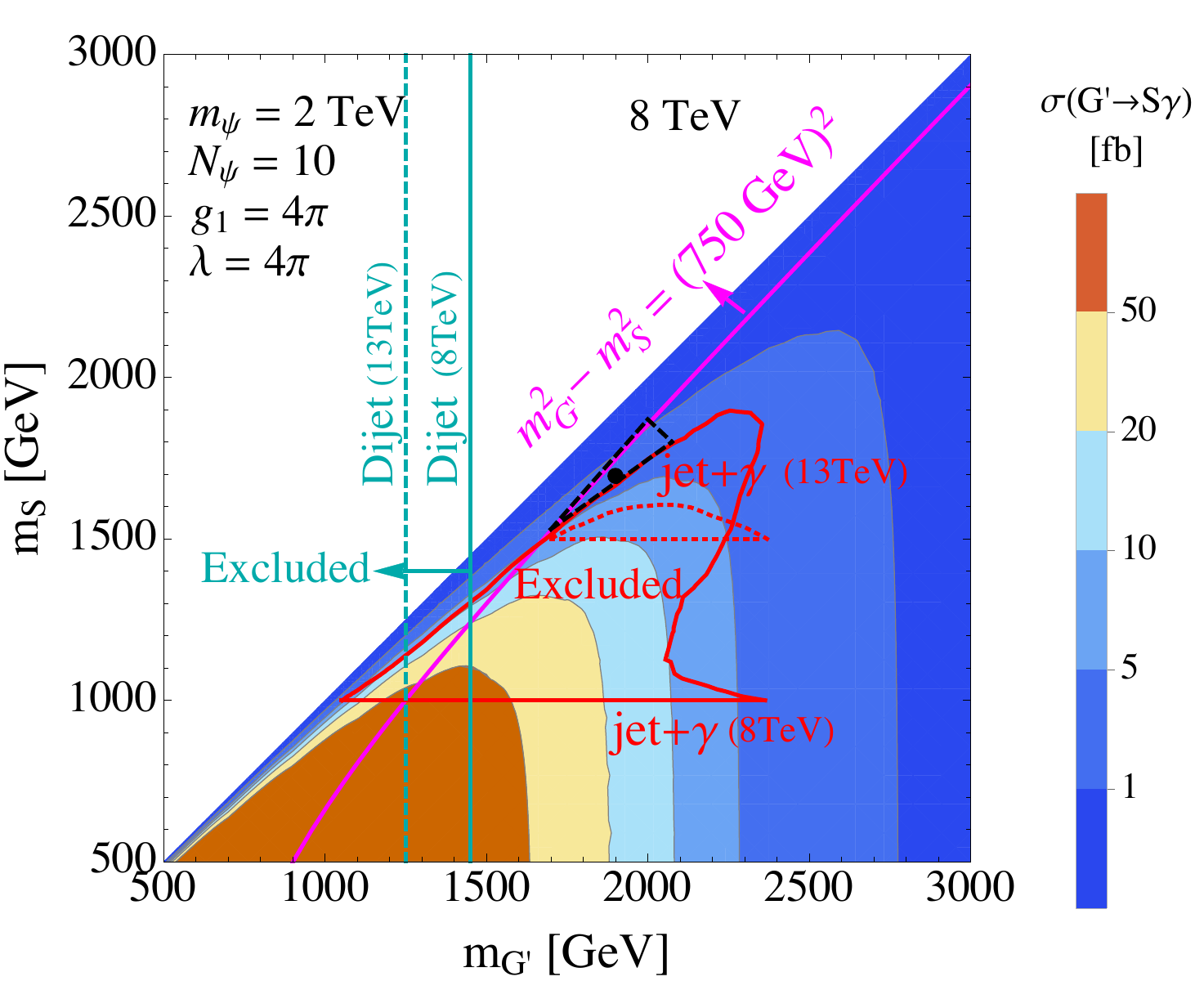}%
    \includegraphics[width=0.5\textwidth]{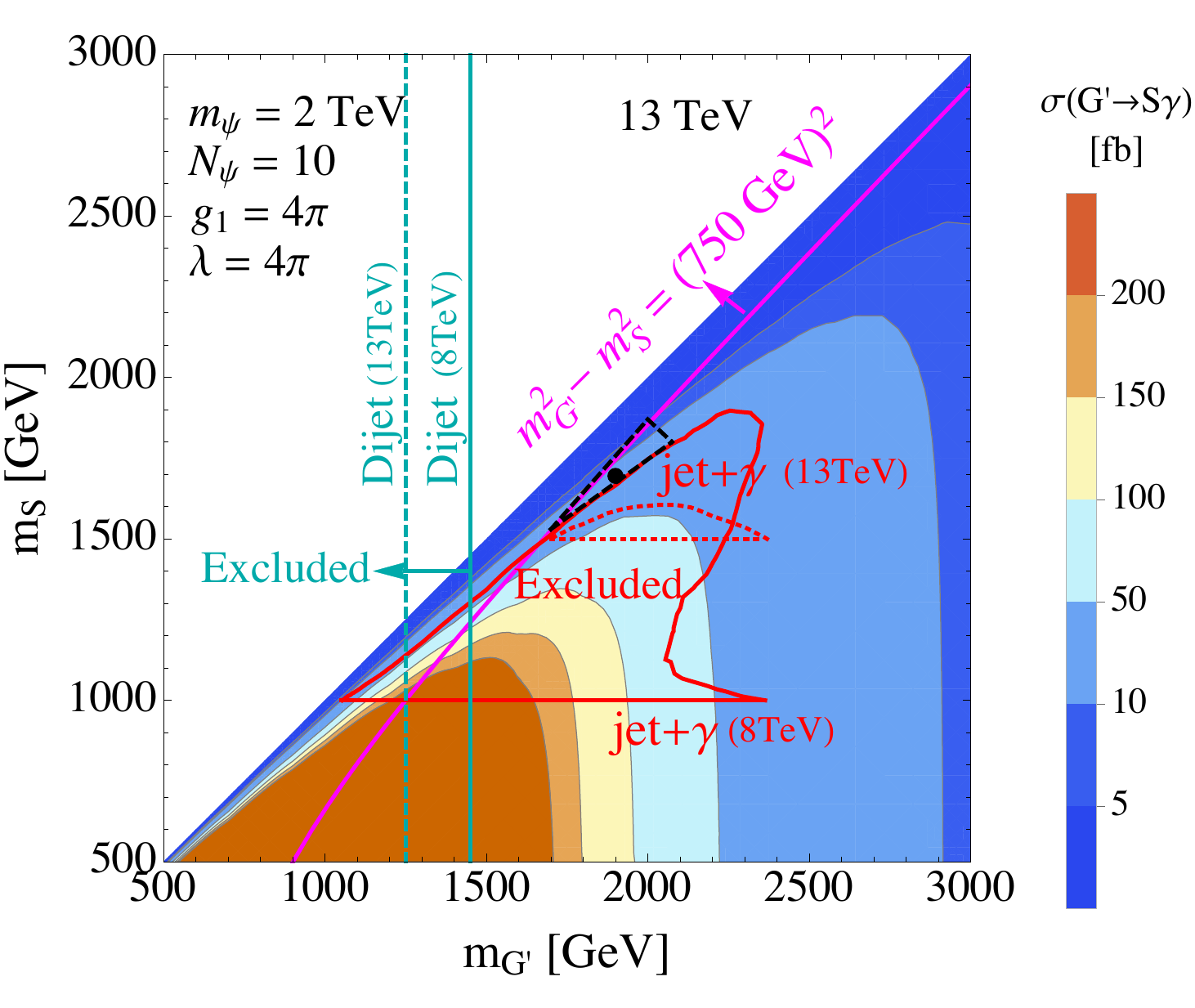}
  \caption{The cross-section for the process $\bar q q \to G' \to
S + \gamma$ for 8 (13) TeV respectively.
The constraints from dijet, jet$+\gamma$ resonance searches are shown for
8 (13) TeV respectively. The requirement from $m_{\gamma\gamma}$ edge in
\cref{eq:m2gammaedge} is shown as magenta line.
The benchmark point is shown as a black dot.
The best fit region for the signal is inside the black dashed triangle.
  }
  \label{fig:constraints-8-13}
\end{figure*}

In summary, the important production of $S$ particles are pair production process
$GG \to SS$
and cascade production $\bar q q \to G' \to S+\gamma$. We add the two cross-section
together and conservatively assume that its acceptance in jet$+\gamma$ resonance
search is $100\%$. In 8 (13) TeV data, we use the $95\%$ CL limits from
excited quark $q^*$. Note that the limits on the $q^*$ mass stops at 1 (1.5) TeV
for 8 (13) TeV data respectively. We show the constraints from jet$+\gamma$ resonance searches in \cref{fig:constraints-8-13}.

To summarize all the constraints, we put the dijet, jet$+\gamma$ resonance searches
in \cref{fig:constraints-8-13}. The figure is the $m_{G'}-m_S$ 2D plot, with other parameters fixed for the signal. The cross-section for the process $\bar q q \to G' \to
S + \gamma$ is given as color code in the plots, for 8 (13) TeV respectively.
The dijets constraints require $G'$ mass should be higher than 1450 (1250) GeV
for 8 (13) TeV data respectively, and is shown as vertical solid(dashed) cyan lines.
The jet$+\gamma$ constraints are shown as solid(dashed) red contours
for 8 (13) TeV data respectively. The region inside the red contours are excluded.
The regions where $m_S$ is smaller than the horizontal red lines are not excluded,
due to lack of data from experiments.

\section{Diphoton signal from cascade decay of $G'$}
\label{sec:diphoton}

In this section, we are going to explain the possible excess in the diphoton events,
via resonant produced $G'$ cascade decay. The decay chain is
$G' \to S+\gamma \to G+\gamma+\gamma$, and the Feynman diagram is given in
\cref{fig:feynman-Gp-resonance}.

Such decay chain has a well-known kinetic edge for the invariant mass of two photons
$m_{\gamma\gamma}$~\cite{Miller:2005zp}.
\begin{align}
\left( {m_{\gamma \gamma }^2 } \right)_{\max }  = \frac{{\left( {m_{G'}^2  - m_S^2 } \right)\left( {m_S^2  - m_G^2 } \right)}}{{m_S^2 }} = m_{G'}^2  - m_S^2
\label{eq:m2gammaedge}
\end{align}

We show our benchmark points $\left\{m_{G'}, m_S\right\} = \left\{1900, 1695\right\}$ GeV in
\cref{fig:constraints-8-13} and found it is consistent with all the dijet and
jet$+\gamma$ resonance constraint. The parameter region which fits the diphoton signal
for both shape and signal strength should have two requirements. First, it should be
close to the $m_{G'}^2  - m_S^2 = (750\text{GeV})^2$ line, which is the requirement
from $m_{\gamma\gamma}$ shape has the correct falling edge around 750 GeV.
Second, it should have a cross-section for $\bar q q \to G' \to S + \gamma$ times
$BR(S\to G +\gamma)=77\%$ in the range about $\mathcal{O}(10)$ fb. This estimation
is coming from the fact that for either short ($\Gamma_{\gamma\gamma}=5$ GeV) or broad
($\Gamma_{\gamma\gamma}=45$ GeV) diphoton resonance, the $2\sigma$ region for signal
cross-section is between $\left[1,11 \right] $ fb~\cite{Falkowski:2015swt}. Our
fit to the signal is not resonance but kinetic edge, thus our signal cross-section
should be larger.
The estimated signal region is given inside the black dashed triangle in
\cref{fig:constraints-8-13}. For the region above the $m_{G'}^2  - m_S^2 = (750\text{GeV})^2$ line, the $m_{\gamma\gamma}$ kinetic edge will move to smaller
values.

Note that when we estimate the constraint, we conservatively
assume signal acceptance equals to $100\%$. The actual limits might be weaker than our conservative limits, thus more parameter space opens up.

\begin{figure*}
    \includegraphics[width=0.6\textwidth]{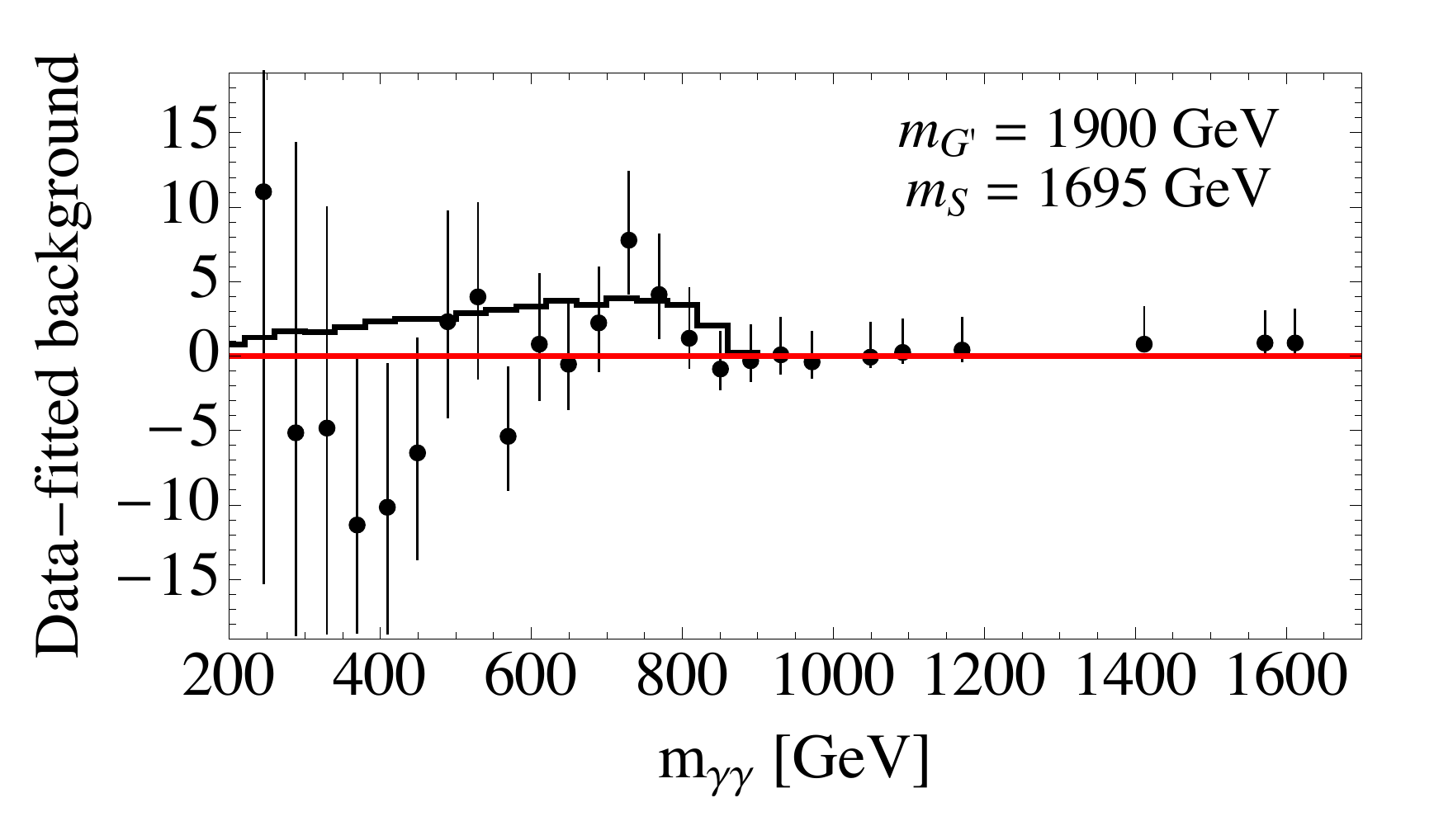}
  \caption{The $m_{\gamma\gamma}$ distribution for the
benchmark point $\left\{m_{G'}, m_S\right\} = \left\{1900, 1695\right\}$ GeV.
The data comes from ATLAS 13 TeV diphoton search \cite{ATLAS-CONF-2015-081}.
The cross-section of the benchmark point is
$\sigma(\bar q q \to G' \to S + \gamma) \times BR(S\to G +\gamma)=14 \rm fb$,
which determines the normalization of the signal.  }
  \label{fig:ATLAS-diphoton-fit}
\end{figure*}

In \cref{fig:ATLAS-diphoton-fit}, we show the $m_{\gamma\gamma}$ distribution for the
benchmark point $\left\{m_{G'}, m_S\right\} = \left\{1900, 1695\right\}$ GeV and compare
with the ATLAS 13 TeV diphoton data. We see that the signal shape provide a right drop
off around 750 GeV. In the small $m_{\gamma\gamma}$ region, our signal provides
significant events there, which is one difference from the real diphoton resonance.
We should remind that in \cref{fig:ATLAS-diphoton-fit}, the normalization of the signal
is not arbitrary but fixed by the cross-section of signal:
$\sigma(\bar q q \to G' \to S + \gamma) \times BR(S\to G +\gamma)=14 \rm fb$.

\begin{figure*}
  \begin{tabular}{cc}
    \includegraphics[width=0.45\textwidth]{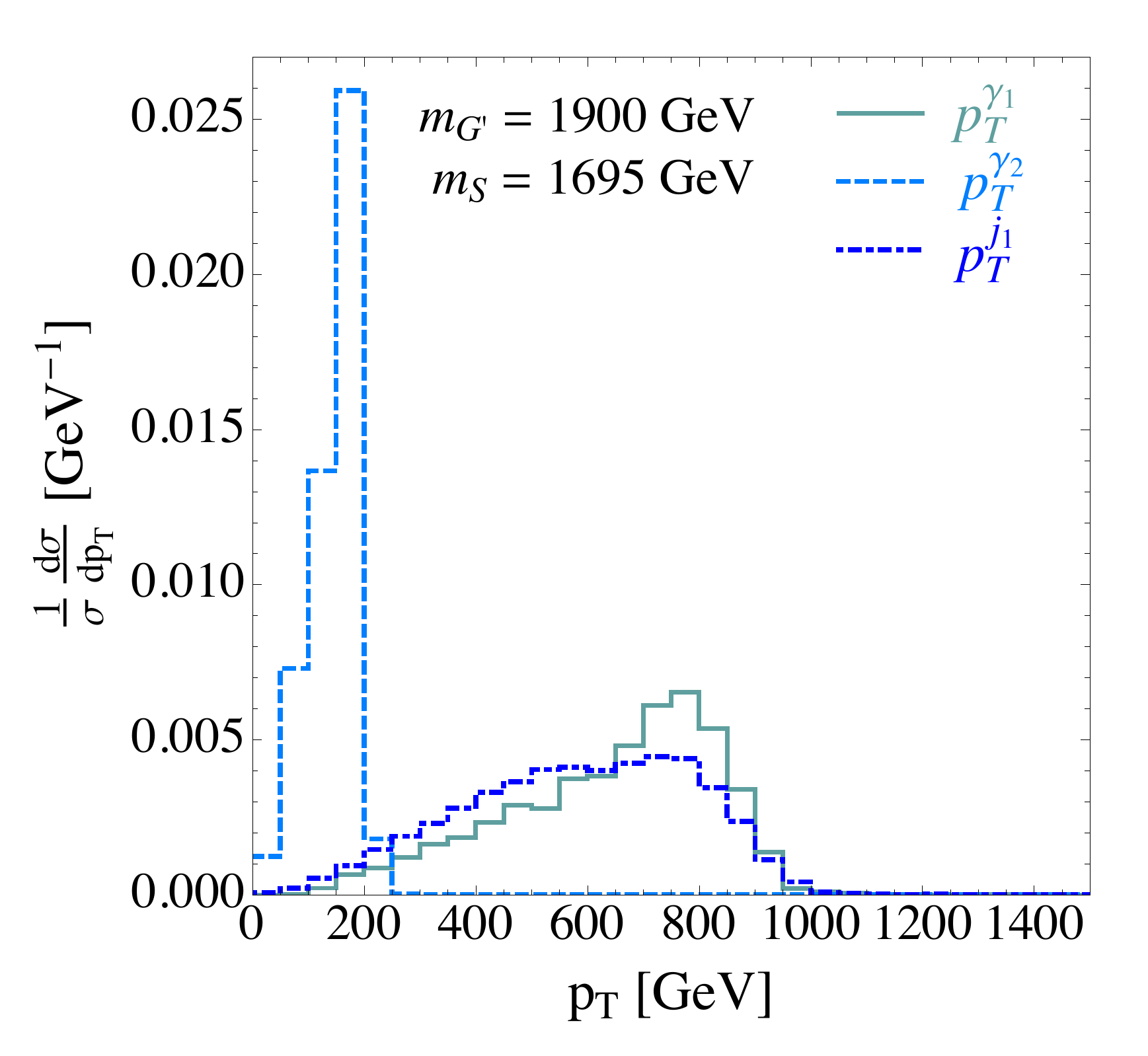} &
    \includegraphics[width=0.45\textwidth]{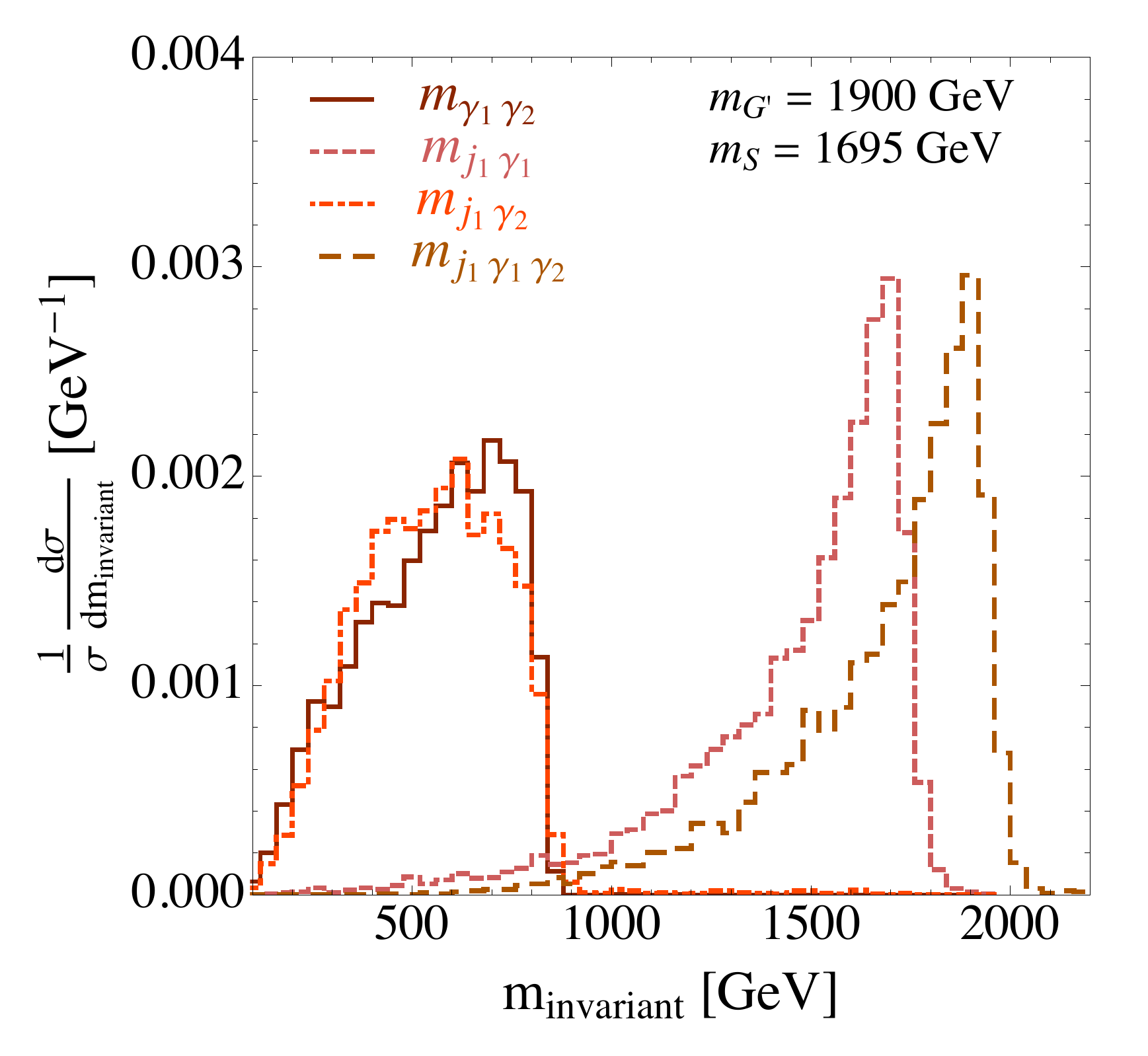}\\
    (a) & (b)
  \end{tabular}
  \caption{The normalized kinetic variable distribution for the benchmark point
  $\left\{m_{G'}, m_S\right\} = \left\{1900, 1695\right\}$ GeV.
  (a) the $p_T$ distribution for leading ($\gamma_1$), subleading
  ($\gamma_2$) photons and leading jet ($j_1$).
  (b) the invariant mass distribution for the combination of $m_{\gamma_1 \gamma_2}$,
  $m_{j_1, \gamma_1}$, $m_{j_1 \gamma_2}$ and $m_{j_1\gamma_1 \gamma_2}$.
  }
  \label{fig:kinematic-distribution}
\end{figure*}

Despite the different $m_{\gamma\gamma}$ shape, our signal also have other
features. The normalized kinetic variable distribution for the benchmark point
$\left\{m_G', m_S\right\} = \left\{1900, 1695\right\}$ GeV are shown in
\cref{fig:kinematic-distribution}.
The left panel shows the $p_T$ distribution for leading ($\gamma_1$), subleading
($\gamma_2$) photons and leading jet ($j_1$). We see that the subleading photon
has $p_T$ much smaller than leading jet, satisfying $p^{\gamma_2}_T < E^{\gamma_2}
\sim (m_{G'}^2 - m_{S}^2)/(2 m_{G'})$. $G'$ is assumed to be produced
at rest in the lab frame, which is a good estimation because it is very heavy.
The leading photon and leading jet has quite large $p_T$, which are also the feature
of our signal. Generally, in diphoton signal from gluon gluon fusion, the leading jet
$p_T$ will be much smaller than our signal.

The right panel shows
the invariant mass distribution for the combination of $m_{\gamma_1 \gamma_2}$,
$m_{j_1, \gamma_1}$, $m_{j_1 \gamma_2}$ and $m_{j_1\gamma_1 \gamma_2}$. We can see
quite easily that $m_{j_1\gamma_1 \gamma_2}$ and $m_{j_1, \gamma_1}$ distribution
show the resonance from $G'$ and $S$. $m_{\gamma_1 \gamma_2}$ distribution drops
around 750 GeV giving an non-resonant interpretation to the possible diphoton excess.
The $m_{j_1 \gamma_2}$ distribution follows $m_{\gamma_1 \gamma_2}$, because $\gamma_1$
is most likely coming from $S$ decay that it has a similar momentum as $j_1$.

In summary, given the above difference in $p_T$ distribution and more invariant mass resonances,
we can distinguish our signal from the other models which directly having a singlet
750 GeV particle decay to diphoton. We eagerly wait for more data to see if
the diphoton excess is a statistic fluctuation or not, and check whether we can
see more resonances in the events to validate our model.

\section{Summary and Conclusion}
\label{sec:conclusions}
In summary, we discuss that the cascade decay of a color octet vector $G'$
at $\mathcal{O}(\rm{TeV})$ can explain
the 750 GeV diphoton excess at LHC 13 TeV. In the coloron model, we also introduce a color octet scalar $S$
coupling to photon via heavy fermion $\Psi$ loop, which leads to the cascade decay
of $G'$. The final states are two photons and one gluon, where the invariant
mass of the two photons has a kinetic edge around 750 GeV, resulting in the diphoton excess.

We start with the $\rm{SU}(3)_1 \times \rm{SU}(3)_2 \rightarrow \rm{SU}(3)_{\rm C}$
renormalizable model, which helps us understanding the relation between different
couplings as discussed in \cref{sec:model}. This setup
reduces the number of free parameters and helps to find the right parameter region to interpret the diphoton excess. Generally, the resonance
production of $q \bar q \to G'$ followed by cascade decay,
faces the diphoton constraint from 8 TeV search.
However in our case, the mass of $G'$ is around TeV scale which is larger than $750$ GeV, so that the quark luminosity function ratio can be quite large between 13 TeV and 8 TeV.
Finally, we also check the collider constraints on this model, including resonance
searches in dijet, $t \bar t $ and jet$+\gamma$ channels at both 8 (13) TeV as shown in \cref{fig:constraints-8-13}.
Our benchmark point for the diphoton excess is safe under these constraints.

To conclude, the diphoton excess can be explained by
the renormalizable coloron model, which has a number of unique features.
 First, all the new particles are charged under $\rm{SU}(3)_{\rm C}$, either triplet or octet.
Secondly, unlike the diphoton decay from a scalar particle, we explain the diphoton
excess as the kinetic edge of cascade decay. Thirdly, we have two new resonances
in the event. One is a jet$+\gamma$ resonance from octet scalar, and the other is
jet$+2\gamma$ resonance from octet vector. Finally, the signal event usually contains
a subleading photon with $p_T \lesssim 200 $ GeV photon and a quite hard leading jet.
With the above unique features, the model is quite easy to be tested with more
data from LHC.

\section*{Acknowledgments}
We thank Michael J. Baker, Joachim Kopp, Ninetta Saviano, Yotam Soreq, Andrea Thamm,
Jesse Thahler, Maikel de Vries, Felix Yu and Hua Xing Zhu for useful discussions.
The work of JL is supported by the German Research Foundation (DFG)
in the framework of the Research Unit ``New Physics at the Large Hadron
Collider'' (FOR~2239) and of Grant No.\ \mbox{KO~4820/1--1}.
The work of WX is support by the U.S. Department of
Energy under grant Contract Numbers DE-SC00012567 and DE-SC0013999.

\bibliographystyle{JHEP}
\bibliography{referencelist}

\end{document}